\documentclass[twocolumn]{jpsj3_reviced}
\usepackage{graphicx}
\usepackage{dcolumn}
\usepackage{bm}
\usepackage{color}%

\def\up{\uparrow}
\def\down{\downarrow }
\def\Vec#1{\bm{#1}}

\newcommand{\imu}{{\rm i}}

\newcommand{\rmd}{{\rm d}}

\title{Numerical construction of a low-energy effective Hamiltonian in a self-consistent Bogoliubov-de Gennes approach of superconductivity}

\author{Yuki \surname{Nagai}$^{1}$, Yasushi \surname{Shinohara}$^{2,3}$, Yasunori \surname{Futamura}$^{4}$, Yukihiro \surname{Ota}$^{5}$, and 
Tetsuya \surname{Sakurai}$^{4,6}$
}
\inst{
\address{$^{1}$CCSE, Japan  Atomic Energy Agency, 5-1-5 Kashiwanoha, Kashiwa, Chiba 277-8587, Japan} \\
$^{2}$Faculty of Pure and Applied Sciences, University of Tsukuba, Tsukuba, Ibaraki 305-8571, Japan \\
$^{3}$Max-Planck-Institut f\"{u}r Mikrostrukturphysik, Weinberg 2, D-06112
Halle, Germany \\
$^{4}$Department of Computer Science, University of Tsukuba, Tsukuba, Ibaraki 305-8573, Japan \\
$^{5}$CEMS, RIKEN, Wako-shi, Saitama 351-0198, Japan \\
$^{6}$JST, CREST, 5, Sanbancho, Chiyoda-ku, Tokyo 102-0075, Japan
}
\abst{
We propose a fast and efficient approach for solving the Bogoliubov-de
 Gennes (BdG) equations in superconductivity, with a numerical matrix-size
 reduction procedure proposed by Sakurai and Sugiura [J. Comput. Appl. Math. 
{\bf 159} (2003) 119]. 
The resultant small-size Hamiltonian contains the information of the original BdG
 Hamiltonian in a given low-energy domain. 
Thus, the present approach leads to a numerical construction
 of a low-energy effective theory in superconductivity. 
The combination with the polynomial expansion method allows a
 self-consistent calculation of the BdG equations. 
Through numerical calculations of quasi-particle excitations in a vortex
 lattice, thermal conductivity, and nuclear magnetic relaxation rate, we
 show that our approach is suitable for evaluating physical quantities
 in a large-size superconductor and a nano-scale superconducting device,
 with the mean-field superconducting theory. 
}

\kword{superconductivity, Bogoliubov-de Gennes equations, Eigenvalue solver}

\begin{document}
\maketitle

\section{Introduction}
To solve an eigenvalue equation is one of the central issues in condensed
matter physics. 
The ground state in a many-body system is nothing but the eigenvector
associated with the lowest eigenvalue of a many-body Hamiltonian. 
The Lanczos algorithm in the exact diagonalization\,\cite{Dagotto:1994}
is suitable for this issue. 
The critical temperature in superconductivity is evaluated by the
greatest eigenvalue of the linearized Eliashberg equations. 
The power iteration algorithm is useful for solving these equations. 
Thus, a lot of efficient methods for either minimum or maximum
eigenvalues have been developed. 

In superconductors, low-energy quasiparticle excitations are quite
important for examining thermodynamic quantities, transport
properties, and so on. 
Their energy scale is characterized by a superconducting gap energy
\mbox{($\sim {\rm meV}$)}, much smaller than a band width 
\mbox{($\sim {\rm eV}$)}. 
In the mean-field Bardeen-Cooper-Schrieffer (BCS) theory, these
excitations correspond to the eigenvalues at the center of an energy
distribution of the Bogoliubov-de Gennes (BdG) Hamiltonian\cite{degennes}.  
This is a direct consequence of the particle-hole symmetry of the BdG
Hamiltonian. 
Furthermore, such an {\it intermediate} region can include zero
eigenvalues (i.e., zero modes). 
The zero modes are related to fundamental properties of
topological insulators and superconductors\,\cite{Teo;Kane:2010}.  
Therefore, in order to study bulk properties of various superconductors
and nano-scale devices with topological materials from atomic-scale physics, an
efficient method to obtain an intermediate spectral region of the BdG
Hamiltonian is highly desirable. 

Typically, the full diagonalization method is used for solving the BdG
equations\,\cite{Takigawa,Takigawa;Machida:2002,Andersen;Schmid:2007,Andersen;Hirshfeld:2008}.  
However, this approach requires a lot of computational memories and a
long computational time. 
In contrast, the polynomial expansion
method\,\cite{Covaci,Zha,Han,NagaiJPSJ,NagaiQPI} allows efficient
self-consistent calculations in superconductivity, without any
diagonalization.   
This approach drastically reduces a computational cost and has an
excellent parallel efficiency, but does not 
lead to direct calculations of eigen-pairs (eigenvalues and eigenvectors) of the 
BdG Hamiltonian.
Thus, this method is not suitable for calculating
dynamical correlation functions (two-particle Green's
functions), which lead to important quantities such as spin/charge
susceptibilities, nuclear magnetic relaxation rate, optical/thermal
conductivities.   
Hence, the algorithms for treating an intermediate energy region of the
BdG Hamiltonian have not been adequately studied.

In this paper, we propose a fast and efficient method for numerically
calculating the eigenvalues and the eigenvectors of the BdG equations.  
Our approach is the combination of the polynomial expansion method with
a contour-integral-based method developed by one of the present authors
(TS) and Sugiura (Sakurai-Sugiura method)\cite{SS,TSakurai,Futamura,Maeda}. 
The Sakurai-Sugiura (SS) method allows us to extract
the eigen-pairs whose eigenvalues are located in
a given domain on the complex plane, from a generic
matrix. Therefore, setting this domain around the
origin of $\mathbb{C}$, an effective Hamiltonian
for the full BdG Hamiltonian can be constructed,
with keeping the information relevant to low-energy
excitations in a superconductor.
A contour-integral representation of the projection operator onto an 
energy domain plays a crucial role.  
We obtain an effective Hamiltonian only with a gapless surface
state in topological insulators and superconductors in large-scale
systems, for example. 

Let us summarize our approach for calculating physical quantities in the
mean-field superconducting theory.  
First, we perform a self-consistent calculation of the BdG equations to
obtain a superconducting gap function. 
Next, we numerically derive a low-energy (small-size) effective
Hamiltonian from the BdG Hamiltonian with the resultant
superconducting gap.
Finally, we calculate physical quantities using the eigenvalues and the
eigenvectors of this effective Hamiltonian. 

This paper is organized as follows. 
In Sec.~\ref{sec:formulation}, we show a general formulation of a
mean-field fermionic theory. 
The Green's functions are expressed by the eigenvalues
and the eigenvectors of the BdG equations. 
In Sec.~\ref{sec:polynomial_expansion}, we explain the
polynomial expansion scheme. 
The application of the SS method to superconductivity is proposed in
Sec.~\ref{sec:SS_method}.  
We show the theoretical background of this approach and the algorithm. 
We stress that the present algorithm is suitable for parallel
computation since the procedure is composed of solving a set of
the linear equations which are independent of each other. 
In Sec.~\ref{sec:demonstrations}, we show the results for typical
examples, as well as the computational costs of the present proposal. 
We perform large-scale calculations in various physical situations. 
As for inhomogeneous systems, we consider a vortex lattice on a
two-dimensional square lattice. 
The quasiparticle excitation spectrum is obtained, varying a magnetic
field and a coherence length.  
The thermal conductivity is also evaluated. 
Moreover, we examine temperature-dependence of the nuclear magnetic
relaxation rate in an $s$-wave superconductor, as an example of a
uniform superconductor. 
These demonstrations indicate that the present approach is a fast and
accurate method for numerically constructing a low-energy effective
Hamiltonian in the mean-field superconducting theory. 
Section \ref{sec:conclusion} is devoted to the summary.

\section{Formulation}
\label{sec:formulation}
\subsection{Hamiltonian}
\label{subsec:hamiltonian}
Throughout this paper, we set $\hbar=k_{\rm B}=1$. 
Let us consider a Hamiltonian for a fermionic many-body
system, 
\begin{equation}
 H 
= 
\frac{1}{2}\psi^{\dagger}\hat{H}\psi
=
\frac{1}{2}
(\bar{c}^{\rm T},c^{\rm T})
\left(
\begin{array}{cc}
\hat{A}         &  \hat{B} \\
-\hat{B}^{\ast} & -\hat{A}^{\ast}
\end{array}
\right)
\left(
\begin{array}{c}
c \\
\bar{c}
\end{array}
\right),
\end{equation}
with 
\mbox{
\(
c=(c_{1},c_{2},\ldots,c_{N})^{\rm T}
\)} 
and 
\mbox{
\(
\bar{c}=(c_{1}^{\dagger},c_{2}^{\dagger},\ldots,c_{N}^{\dagger})^{\rm T}
\)}. 
Here, the symbol ${\rm T}$ represents transposition. 
The fermionic annihilation and creation operators are denoted as,
respectively, $c_{i}$ and $c_{i}^{\dagger}$ ($i=1,\ldots,N$). 
The index $i$ includes all the relevant degrees of freedom such as
spatial sites, spins, orbitals, and so on. 
The canonical anti-commutation relation is 
\mbox{$[c_{i},c_{j}^{\dagger}]_{+}=\delta_{ij}$}. 
The Hamiltonian matrix $\hat{H}$ is a $2N\times 2N$ Hermite matrix. 
The hermitian property of $H$ and the canonical anti-commutation
relation imply that the $N\times N$ complex matrices $\hat{A}$ and
$\hat{B}$ in $\hat{H}$ satisfy 
\(
\hat{A}^{\dagger} = \hat{A}
\)
and 
\(
\hat{B}^{\rm T} = -\hat{B}
\). 
In the case of superconductivity, $\hat{H}$ corresponds to the
mean-field BCS Hamiltonian and $\hat{B}$ contains superconducting gaps. 

\subsection{Bogoliubov-de Gennes equations}
The BdG equations are regarded as the eigenvalue equations with respect
to $\hat{H}$ 
\begin{align}
\hat{H} \Vec{x}_{\gamma} 
&= \epsilon_{\gamma} \Vec{x}_{\gamma}, \: \:
(\gamma = 1,2,\cdots,2 N),  
\end{align}
with $\Vec{x}_{\gamma} = (\Vec{u}_{\gamma},\Vec{v}_{\gamma})^{\rm T}$. 
The column vectors $\Vec{u}_{\gamma}$ and $\Vec{v}_{\gamma}$ are
$N$-component complex vectors.  
To solve the BdG equations is equivalent to the diagonalization of
$\hat{H}$ with a unitary matrix $\hat{U}$.
The matrix elements of $\hat{U}$ are 
\begin{align}
U_{i,\gamma} &= u_{\gamma,i}, \: \: U_{i+N,\gamma} = v_{\gamma,i}.
\end{align}
The eigenvalues $\epsilon_{\gamma}$ are not independent of each other. 
In fact, using the {\it particle-hole}
transformation\,\cite{Hemmen:1980} such that
\mbox{
\(
\hat{J}\Vec{x}_{\gamma} 
= (\Vec{v}_{\gamma}^{\ast},\,\Vec{u}_{\gamma}^{\ast})^{\rm T}
\)
}, one can show that 
\mbox{
\(
\hat{H} (\hat{J}\Vec{x}_{\gamma}) 
= - \epsilon_{\gamma} (\hat{J}\Vec{x}_{\gamma})
\)}. 

\subsection{Two-particle Green's functions}
Two-particle Green's functions are related to different
physical quantities in condensed matter physics. 
Here,they are written in terms of the solutions of the BdG
equations.  
Let us consider a two-particle Green's function in the imaginary time
$\tau$, 
\begin{subequations} 
\begin{align}
Q_{1234}(\tau) &= \langle 
{\rm T}_{\tau} [ 
c_{i_{1}}^{\dagger}(\tau) c_{i_{2}}(\tau) 
c_{i_{3}}^{\dagger}(0) c_{i_{4}}(0)
]
\rangle, \\
&= 
G_{i_{2} i_{3}}(\tau) \bar{G}_{i_{1} i_{4}}(\tau) 
- 
F_{i_{2} i_{4}}(\tau) \bar{F}_{i_{1} i_{3}}(\tau),
\end{align}
\end{subequations} 
with the one-particle Green's functions 
\begin{subequations}
\begin{align}
G_{i j}(\tau) &= - \langle 
{\rm T}_{\tau} [ c_{i}(\tau)c_{j}^{\dagger}(0) ] \rangle, \\
F_{i j}(\tau) &= - \langle 
{\rm T}_{\tau} [ c_{i}(\tau)c_{j}(0) ] \rangle, \\
\bar{F}_{i j}(\tau) &= - \langle 
{\rm T}_{\tau} [ c_{i}^{\dagger}(\tau)c_{j}^{\dagger}(0) ] \rangle, \\
\bar{G}_{i j}(\tau) &= - \langle 
{\rm T}_{\tau} [ c_{i}^{\dagger}(\tau)c_{j}(0) ] \rangle.
\end{align}
\end{subequations} 
With the use of the relation
\begin{align}
\int_{0}^{\beta} d \tau e^{\imu \Omega_{m} \tau} A(\tau) B(\tau) &=
 \frac{1}{\beta} \sum_{\omega_{n}} A(\imu \omega_{n}) 
B(\imu \Omega_{m} - \imu \omega_{n}),
\end{align}
the Fourier transformed function $Q(\imu \Omega_{m})$ is 
\begin{align}
Q_{1234}(\imu \Omega_{m}) 
&= \frac{1}{\beta} \sum_{\omega_{n}} \left[
G_{i_{2} i_{3}}(\imu \omega_{n}) 
\bar{G}_{i_{1} i_{4}}(\imu \Omega_{m} - \imu \omega_{n}) \right. 
\nonumber \\
&\left. 
- F_{i_{2} i_{4}}(\imu \omega_{n}) 
\bar{F}_{i_{1} i_{3}}(\imu \Omega_{m} - \imu \omega_{n}) 
\right]. \label{eq:q1234}
\end{align}
Here, $\beta$ is the inverse temperature and $\omega_{n} = \pi (2
n + 1)/\beta$ and $\Omega_{m} = \pi (2 m)/\beta$ are the fermionic and
bosonic Matsubara frequencies, respectively. 
The one-particle Green's functions are written as a $2N\times 2N$
matrix, 
\begin{equation}
\hat{G}(z) 
=
\int_{-\infty}^{\infty}
\frac{d\omega}{2\pi}
\frac{\hat{A}(z)}{z-\omega},
\quad
\hat{A}_{\alpha \beta}(\omega)
= \sum_{\gamma=1}^{2 N} U_{\alpha, \gamma} U_{\beta, \gamma}^{\ast} 
\delta(\omega-\epsilon_{\gamma}).
\end{equation}
We find that 
$G_{ij} = \hat{G}_{ij}$, 
$F_{ij} = \hat{G}_{i, j+N}$, 
$\bar{F}_{ij} = \hat{G}_{i + N, j}$, 
and $\bar{G}_{ij} = \hat{G}_{i+N, j+N}$. 
We can phenomenologically describe a dissipation effect, replacing the
delta function in $\hat{A}$ with an approximate $\delta$-function.  
The dynamical correlation function with the real energy $\Omega$ is 
\begin{align}
Q_{1234}(\Omega) 
&= 
\sum_{\gamma,\gamma'}^{2 N} 
U_{i_{2}, \gamma} U_{i_{1} + N, \gamma'}
\left[ 
U_{i_{3}, \gamma}^{\ast} U_{i_{4}+N, \gamma'}^{\ast} \right. \nonumber \\
&\left.  - U_{i_{4}+N, \gamma}^{\ast} U_{i_{3}, \gamma'}^{\ast} \right] 
 \frac{f(\epsilon_{\gamma}) - f(-\epsilon_{\gamma'})}{\Omega + i \eta - (\epsilon_{\gamma} + \epsilon_{\gamma'})}, \label{eq:dynam}
\end{align}
with setting $\imu \Omega_{m} \rightarrow \Omega + i \eta$ 
($\eta \rightarrow 0^{+}$)  in Eq.~(\ref{eq:q1234}). 
Here, $f(x) = 1/(e^{\beta x} + 1)$ denotes the fermion distribution
function. 

\section{Polynomial expansion method}
\label{sec:polynomial_expansion}
We briefly summarize the polynomial expansion method 
for a self-consistent calculation of the BdG equations, according to our
previous paper\cite{NagaiJPSJ}. 
The essence is the expansion of the spectral density of the Green's
functions, with orthonormal polynomials in $[-1,1]$
satisfying 
\begin{subequations}
\begin{align}
&
 \delta(x-x^{\prime}) 
= \sum_{n=0}^{\infty}\frac{W(x)}{w_{n}}
\phi_{n}(x)\phi_{n}(x^{\prime}), 
\label{eq:delta}
\\
&
w_{n}\delta_{n,m} = 
\int_{-1}^{1} \phi_{n}(x)\phi_{m}(x) W(x) dx, \label{eq:wn} 
\\
&
\phi_{n+1}(x) = (a_{n}+b_{n}x)\phi_{n}(x)  -
 c_{n}\phi_{n-1}(x). \label{eq:rec} 
\end{align}
\end{subequations} 
The Chebyshev polynomial is often used, because the resultant
formulae become the simplest forms. 
The application of the other polynomials is discussed in, e.g., our previous
paper\,\cite{Nagaipro}.

The spectral density (matrix) is given as a difference between the retarded and
the advanced Green's functions, 
\mbox{
\(
\hat{d}(\omega) 
= \hat{G}(\omega + i0) - \hat{G}(\omega - i0)\)}. 
Let us expand $\hat{d}(\omega)$ by $\phi_{n}(x)$, rescaling
$\hat{H}$ and $\omega$ so that $\hat{\mathcal{K}}=(\hat{H}-b)/a$ and 
$x=(\omega - b)/a$, with 
$a=(E_{\rm max}-E_{\rm min})/2$ 
and 
$b=(E_{\rm max}+E_{\rm min})/2$. 
Here, $E_{\rm max}$ and $E_{\rm min}$ are energy scales satisfying
$E_{\rm min} \le \epsilon_{\gamma} \le E_{\rm max}$. 
The elements of $\hat{d}(\omega)$ are related to various correlation
functions. 
Using the constant vectors $\Vec{e}(i)$ and $\Vec{h}(i)$ such that
$[\Vec{e}(i)]_{\gamma} = \delta_{i,\gamma}$ and  
$ [\Vec{h}(i)]_{\gamma} = \delta_{i+N,\gamma}$, 
we obtain 
\begin{subequations} 
\begin{align}
\langle c^{\dagger}_{i} c_{j} \rangle &= 
\sum_{n = 0}^{\infty}
\Vec{e}(j)^{\rm T} \Vec{e}_{n}(i) \frac{{\cal T}_{n}}{w_{n}}, \label{eq:cdct}\\
\langle c_{i} c_{j} \rangle &= 
\sum_{n = 0}^{\infty}
\Vec{e}(j)^{\rm T} \Vec{h}_{n}(i) \frac{{\cal T}_{n}}{w_{n}}, \label{eq:cct} 
\end{align}
\end{subequations} 
where 
\begin{align}
&
{\cal T}_{n} = 
\int_{-1}^{1} \rmd x f(ax + b) W(x) \phi_{n}(x).
\end{align}
A sequence of a vector $\Vec{q}_{n} = \phi_{n}({\cal K})\Vec{q}$
\mbox{[$\Vec{q}=\Vec{e}(i),\,\Vec{h}(i)$]}
is recursively obtained by  
\begin{subequations}
\begin{align}
 \Vec{q}_{n+1} 
&= (a_{n} + b_{n}\hat{\mathcal{K}}) \Vec{q}_{n} -
  c_{n}\Vec{q}_{n-1}
\quad
(n\ge 2), \label{eq:rec_matrix} \\
 \Vec{q}_{1} 
&= \phi_{1}(\hat{\mathcal{K}})\Vec{q}, 
\quad \Vec{q}_{0} 
= \phi_{0}(\hat{\mathcal{K}})\Vec{q}.
\end{align}
\end{subequations} 
The use of the recurrence formula leads to a self-consistent calculation 
of the BdG equations, without any diagonalization of $\hat{H}$. 

\section{Contor-integral-based method (Sakurai-Sugiura method)}
\label{sec:SS_method}
In this paper, we use the SS method for finding eigenvalues in a given
energy domain and their associated eigenvectors. 
This approach is a numerical solver for a generalized
eigenvalue problem so that 
\(
A \Vec{x} = \epsilon B \Vec{x}
\), with $A,\,B \in \mathbb{C}^{n_{\rm s}\times n_{\rm s}}$, 
and has been applied to various physical issues such as the real-space
density functional theory\cite{RSDFT} and the lattice quantum
chromodynamics\cite{QCD}.  
In this paper, $B$ is the identity matrix, and $A$ is an Hermite
matrix. 

Our aim is to reduce the size of $A$, keeping as much information of
the eigenvalues and the eigenvectors as possible. 
Let us consider the use of an $n_{\rm s} \times m_{\rm s}$ 
($n_{\rm s} \ge m_{\rm s}$) matrix $Q$, whose $i$th column is
$\bm{q}_{i} \in \mathbb{C}^{n_{\rm s}}$ 
(i.e., $Q = \{ \bm{q}_{1},\ldots, \bm{q}_{m_{\rm s}}\}$). 
Here $\{\bm{q}_{i}\}_{i=1}^{m_{\rm s}}$ is a set of linearly independent
vectors in $\mathbb{C}^{n_{\rm s}}$. 
We obtain an $m_{\rm s}\times m_{\rm s}$ matrix 
\begin{equation}
 \tilde{A} = Q^{\dagger} A Q. 
\end{equation}
We denote a given energy domain of $A$ as 
$\mathcal{E}(\subset\mathbb{R})$.  
Let us suppose that $\bm{q}_{i}$ is represented by a linear combination
of $\{\bm{x}_{j}\}_{j=1}^{m_{\rm s}}$, where 
$A \bm{x}_{j} = \epsilon_{j} \bm{x}_{j}$ 
($\epsilon_{j} \in \mathcal{E}$). 
Thus, $\tilde{A}$ contains $m_{\rm s}$ eigenvalues of $A$ ($\{ \epsilon_{j} \}_{j=1}^{m_{\rm s}}$). 
It is necessary for implementing this procedure to know parts of
the eigenvectors of $A$. 
Furthermore, one has to carefully choose $m_{\rm s}$ to avoid losing the
relevant information of $A$. 
Remarkably, this issue will be solved by an approximate
evaluation of contour integrals associated with a projection operator
onto eigenspaces of $A$. 
All the steps of the algorithm is summarized in
Sec.\,\ref{subsec:algorithm_ss}. 

\subsection{Projection and moment vectors}
\begin{figure}[thb]
\begin{center}
\includegraphics[width = 0.6\columnwidth]{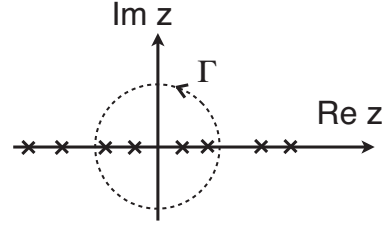}
\caption{\label{fig:fig1}
Schematic diagram of a contour on $\mathbb{C}$.
}
\end{center}
\end{figure}

We start with a way to make a projection onto a target subspace spanned
by $\{\bm{x}_{j}\}_{j=1}^{m_{\rm s}}$ ($||\bm{x}_{j}|| = 1$). 
An arbitrary $n_{\rm s}$-dimensional vector $\bm{v}$ is expanded by
$\{\bm{x}_{i}\}$, with $n_{\rm s}$ complex coefficients, 
\(
\bm{v} = \sum \alpha_{i} \bm{x}_{i}
\). 
We define the projection $P_{\Gamma}(A)$ as 
\begin{equation}
 P_{\Gamma}(A) \bm{v} 
= 
 \sum_{j=1}^{m_{\rm s}}\alpha_{j}\bm{x}_{j}.\label{eq:proj}
\end{equation}
Using $P_{i} = \bm{x}_{i}\bm{x}_{i}^{\dagger}$, we
find that the resolvent\cite{textM} of $A$ is 
\begin{equation}
 \frac{1}{zI - A} 
=
\sum_{i=1}^{n_{\rm s}}\frac{P_{i}}{z-\epsilon_{i}}
\quad
(z \in \mathbb{C}\backslash \sigma(A)),
\end{equation}
with the $n_{\rm s}\times n_{\rm s}$ identity matrix, $I$ and a set of
all the eigenvalues of $A$, $\sigma(A)$, since 
\(
A = \sum \epsilon_{i}P_{i}
\). 
Let us suppose that the $m_{\rm s}$ distinct eigenvalues (i.e., simple
poles on $\mathbb{C}$) are located inside a closed loop $\Gamma$ on
$\mathbb{C}$, and the others are outside $\Gamma$, as shown in
Fig.~\ref{fig:fig1}. 
Thus, we obtain a contour-integral representation of $P_{\Gamma}(A)$ 
\begin{equation}
 P_{\Gamma}(A)
=
\oint_{\Gamma} \frac{dz}{2\pi \imu }
\frac{1}{zI-A}. 
\end{equation}

Now, let us write essential quantities for determining the
reduction matrix $Q$. 
The {\it moment} vector $\bm{s}_{k}$ ($k=0,\,1,\ldots,\,M-1$) is defined as
$\bm{s}_{k}=A^{k}P_{\Gamma}(A)\bm{v}$, with a vector
$\bm{v}\in\mathbb{C}^{n_{\rm s}}$. 
From its contour-integral representation, we find that $\bm{s}_{k}$ is
related to the $k$th moment, 
\begin{equation}
 \bm{s}_{k} 
= 
 \oint_{\Gamma} \frac{dz}{2\pi \imu } \frac{z^{k}}{zI - A} \bm{v}.
\end{equation}
An important property of $\Vec{s}_k$ is that this is a vector in 
a vector space
associated with $P_{\Gamma}(A)$.
This fact is checked by applying $A^k$ to Eq.~(\ref{eq:proj}). 
We remark that all the moment vectors are linearly independent of each
other for an arbitrary $\bm{v}$, since the subspace spanned by
$\{\bm{s}_{k}\}_{k=0}^{M-1}$  is the order-$M$ Krylov
subspace\cite{Watkins:2008} generated by $A$, 
${\cal K}_{M}(A,P_{\Gamma}(A)\Vec{v}) =  {\rm span} \: \{ 
P_{\Gamma}(A) \Vec{v},A P_{\Gamma}(A) \Vec{v}, A^{2} P_{\Gamma}(A)
\Vec{v},  \cdots, A^{M-1} P_{\Gamma}(A) \Vec{v} \}$. 
In our algorithm to determine $Q$, $M$ is an input parameter. 
Then, one has to construct linearly independent vectors from
$\{\bm{s}_{k}\}_{k=0}^{M-1}$, varying $\bm{v}$, and evaluate $m_{\rm s}$ with a
proper manner. 
Another important issue is to numerically calculate the contour integrals. 
These points will be explained in the following.

\subsection{Approximation of contour integrals with numerical quadrature}
\label{subsec:numerical_quadrature}
We show a method to approximate a contour integral with numerical
quadrature.  
Let us suppose that a Jordan curve $\Gamma$ on $\mathbb{C}$ is
represented by scaling and shifting another Jordan curve $\Gamma_{0}$, 
with a scaling factor $\rho$ and a shift $\gamma$. 
Without loss of generality, we assume that $\Gamma_{0}$ encloses the
origin on $\mathbb{C}$. 
Let $\zeta(\theta)$ be a point on $\Gamma_{0}$, with a parameter $\theta$
($0\le \theta \le 2\pi$), and let $z$ on $\Gamma$ be given by 
\(
 z(\theta) = \gamma + \rho \zeta(\theta)
\). 
Then, using $N_{q}$-point quadrature rule, the moment vector is
approximately written by 
\begin{equation}
 \bm{s}_{k} 
\sim 
\frac{1}{N_{q}}\sum_{j=1}^{N_{q}}
\rho w_{j} z_{j}^{k} \bm{y}_{j},
\end{equation}
with $w_{j}=w(\theta_{j})$,
\(
w(\theta) = -i \zeta^{\prime}(\theta)
\), 
\(
z_{j} = \gamma + \rho \zeta_{j}
\), 
and 
\(
\zeta_{j} = \zeta(\theta_{j})
\). 
The vector $\bm{y}_{j}$ is the solution of the linear equation 
$(z_{j}I - A )\bm{y}_{j} = \bm{v}$. 
When all the eigenvalues are located on the real axis, it might be
better to put the quadrature points closer to the real axis,  
\begin{equation}
 z_{j} =
\gamma + \rho(\cos\theta_{j} + i \alpha \sin\theta_{j}),
\quad
\theta_{j} = \frac{2\pi}{N_{\rm q}} \left(
j-\frac{1}{2}
\right), \label{eq:daz}
\end{equation}
with vertical scaling factor $\alpha$ ($0< \alpha \le 1$). 
The quadrature weight is 
\begin{equation}
 w_{j} = \alpha \cos\theta_{j} + i\sin\theta_{j}. \label{eq:daw}
\end{equation}
When $\alpha=1$, $\Gamma_{0}$ is a unit circle. 
Other formulae including contour integrals are also calculated by this
$N_{\rm q}$-point numerical quadrature. 

\subsection{Construction of subspace}
\label{subsec:construction_subspace}
Now, we show a way to determine the subspace size and the corresponding
reduction matrix. 
First, we make a sequence of the moment vectors, varying $\bm{v}$. 
The use of this sequence is essential for constructing the linearly
independent vectors and the subspace. 
We set $L$ complex vectors, $\bm{v}^{i} (\in \mathbb{C}^{n_{\rm s}})$ 
($i=1,\,2,\,\ldots,L$), and make an $n_{\rm s}\times L$ real matrix 
\(
\hat{V}=\{ \bm{v}^{1},\bm{v}^{2},\ldots,\bm{v}^{L} \}
\). 
We call $\hat{V}$ a source matrix. 
Then, we obtain an $n_{\rm s}\times L$ matrix $\hat{S}_{k}$, 
\begin{equation}
 \hat{S}_{k} = \frac{1}{N_{\rm q}}\sum_{j=1}^{N_{\rm q}}\rho w_{j} z_{j}^{k} \hat{Y}_{j} , \label{eq:sky}
\end{equation}
with 
\begin{equation}
  (z_{j}I - A) \hat{Y}_{j} = \hat{V}. \label{eq:lin}
\end{equation}
The $i$th column of $\hat{S}_{k}$ is related to $\bm{v}^{i}$, 
\(
\bm{s}_{k}^{i} 
=
(1/N_{\rm q})\sum_{j}\rho w_{j} z_{j}^{k} \bm{y}_{j}^{i}
\), 
with 
\(
(z_{j}I - A) \bm{y}_{j}^{i} = \bm{v}^{i}
\). 
Each element of $\bm{v}^{i}$ is a uniform random variable in $(-1,1)$. 
It means that we make $A^{k}P_{\Gamma}(A)$ via random sampling, with the
source size $L$. 
The integer parameter $L$ is determined by the
calculation of ${\rm Tr}\,P_{\Gamma}(A)$, as seen below. 

Next, we determine the subspace size. 
A bundle of $\hat{S}_{k}$ leads to 
an $n_{\rm s}\times LM$ matrix
\(
\hat{S} = \{\hat{S}_{0},\,\hat{S}_{1},\ldots, \hat{S}_{M-1}\}
\).
Now, we perform the singular-value decomposition of $\hat{S}$, and
obtain the singular values $\{\sigma_{i}\}$, with 
\(
\sigma_{1} \ge \sigma_{2} \ge \ldots \ge 0
\). 
Then, we find the number of the singular values satisfying 
\(
\sigma_{i}/\sigma_{1} > \delta
\), with a small positive constant $\delta (\sim 10^{-14})$. 
Thus, we have an effective rank (i.e., the number of the
predominant linearly independent vectors) of $\hat{S}$. 
We stress that this rank should be greater than or equal to a prior
value $\tilde{m}_{\rm s}$, which is determined by calculating the trace
of the resolvent matrix (see below). 
We use the resultant effective rank as the dimension of the subspace. 
We remark 
a large source size leads to a highly accurate calculation, but causes an extreme increase in the subspace size. 
To avoid this increase for the high accuracy, 
one can use an iterative
refinement\cite{TSakurai} of a subspace, as seen in Appendix
\ref{app:iterative_refinement}. 

Now, the construction of the reduction matrix to the subspace is
straightforward. 
Using a submatrix composed of the first $m_{\rm s}$ columns of $\hat{S}_{k}$
(e.g., $\hat{S}_{k}(:,\,1:m_{s})$ in terms of Fortran 90) and the Gram-Schmidt
orthonormalization, we obtain an 
$n_{\rm s}\times m_{\rm s}$ matrix $\tilde{Q}$ whose $m_{\rm s}$ column
vectors are orthonormal to each other. 
This is our reduction matrix. 
From the construction manner, one finds that $\tilde{Q}$ contains predominant
$m_{\rm s}$ eigenvectors of $A$. 
Alternatively, one may use a matrix $\hat{U}$ such that 
$\hat{S} = \hat{U} \hat{\Sigma} \hat{W}^{\dagger}$ 
and $\hat{\Sigma}={\rm diag}(\sigma_{1},\sigma_{2},\ldots)$. 
This matrix is automatically obtained when performing the singular-value
decomposition of $\hat{S}$ by using the ZGESVD routine of LAPACK, and the $m_{\rm s}$ column-vectors are
orthogonal to each other. 

The source size $L$ and the moment size $M$ have to be carefully
chosen. 
In particular, the integer $LM$, which is the total number of the moment
vectors to take in a simulation, should be as small as possible for a
few computational costs. 
First, we predict the prior rank $\tilde{m}_{\rm s}$, with the
stochastic estimation method\cite{Futamura,Maeda}. 
We prepare an $n_{\rm s}\times L_{0}$ real matrix $\hat{V}$ whose
elements are either $-1$ or $1$ with equal probability. 
Here, $L_{0}$ is an input parameter. 
The number of the eigenvalues inside $\Gamma$ is 
\mbox{
\(
{\rm Tr}\,P_{\Gamma}(A)
\)}, 
 since 
\begin{align}
{\rm Tr}\,P_{\Gamma}(A) &= 
\sum_{i} \oint_{\Gamma} \frac{dz}{2 \pi \imu} \Vec{x}_{i}^{\dagger} \frac{1}{z I - A} \Vec{x}_{i},  \\
&= \sum_{i} {\rm Res} \: \frac{1}{z  - \epsilon_{i}}. 
\end{align}
The stochastic estimation of ${\rm Tr}\,M$ for an 
$n_{\rm s}\times n_{\rm s}$ matrix is 
$ {\rm Tr} \,M  
\sim 
(1/L_{0})\sum_{i=1}^{L_{0}} (\bm{v}^{i})^{\rm T}M \bm{v}^{i}
$(See, Appendix\,\ref{app:estimation_trace}). 
Thus, using the $N_{\rm q}$-point numerical quadrature, $\tilde{m}_{\rm
s}$ is estimated by 
\begin{equation}
 \tilde{m}_{\rm s} 
= \frac{1}{L_{0}} \sum_{i=1}^{L_{0}}(\bm{v}^{i})^{\rm T}\bm{s}_{0}^{i}. \label{eq:ms}
\end{equation}
Then, the source size $L$ is
\begin{equation}
 L 
= \left[
\frac{\kappa \tilde{m}_{\rm s}}{M}
\right], 
\label{eq:source_size}
\end{equation}
with $\kappa \ge 1$. 
The symbol $[x]$ means the smallest integer greater than $x$. 
The value of $LM$ is larger than
$\tilde{m}_{\rm s}$. 
Thus, the requirement that the rank of $\hat{S}$ is greater than and equal to
$\tilde{m}_{\rm s}$ is automatically satisfied.

\subsection{Algorithm of the SS method}
\label{subsec:algorithm_ss}
Now, we show all the steps of the SS method for calculating the
eigenvalues and the eigenvectors of the BdG Hamiltonian $\hat{H}$. 

(i) Set $\hat{H} \in \mathbb{C}^{N \times N}$ ($n_{\rm s}=N$), $L_{0}$,
$M$, $N_{\rm q}$ and 
\mbox{$\hat{V} = \{\Vec{v}^{1},\cdots,\Vec{v}^{L_{0}} \}$}. 
The elements of the sampling vector $\Vec{v}_{i}$ take either -1 or 1,
with equal probability. 

(ii) Solve Eq.~(\ref{eq:lin}) for $Y_{j}, \: j = 1, \cdots, N_{\rm q}$. 
One can solve these equations separately so that parallel computations
can be easily implemented. 

(iii) Compute Eq.~(\ref{eq:sky}).

(iv) Compute Eq.~(\ref{eq:ms}), and estimate $L$ via Eq.~(\ref{eq:source_size}).
 
(v) Give the elements of $\hat{V} = \{\Vec{v}^{1},\cdots,\Vec{v}^{L} \}$
by random numbers and solve Eq.~(\ref{eq:lin}). 

(vi) Compute Eq.~(\ref{eq:sky}) using the results in (v).

(vii) Perform the singular-value decomposition 
\mbox{
$\hat{U} \hat{\Sigma} \hat{W}^{\dagger} = \{\hat{S}_{0},\cdots,\hat{S}_{M-1} \}$} 
and find $m_{\rm s}$ such  
that \mbox{$|\sigma_{j}|/|\sigma_{1}| \leq \delta$} for $1 \leq j \leq m_{\rm s}$. 

(viii) Obtain a matrix $\tilde{Q}$ from $\tilde{Q} =
\hat{U}(:,1:m_{\rm s})$. 

(ix) Form $\tilde{H} = \tilde{Q}^{\dagger} H \tilde{Q}$.

(x) Compute the eigenvalues $\epsilon_{i}$ and eigenvectors
$\Vec{w}_{i}$ of the matrix $\tilde{H}$.  

(xi) Set $\Vec{x}_{i} = \tilde{Q} \Vec{w}_{i}$. 

If one uses the iterative refinement of a subspace, one adopts either
Eqs.~(\ref{eq:vv}) or (\ref{eq:vc}), and goes to (vi).  
The one- or two- particle Green's function are calculated by the
eigen-pair $(\epsilon_{i},\Vec{x}_{i})$. 

\section{Numerical demonstrations}
\label{sec:demonstrations}
We show the effectiveness and the validity of the present approach,
focusing on a single-band superconductor. 
Hereafter, the index $i$ in Sec.~\ref{subsec:hamiltonian} indicates a
spatial site on a two-dimensional square lattice. 
The spin indices ($\up$ and $\down$) are explicitly written in the
creation and annihilation operators. 
We consider a two-dimensional $L_{x} \times L_{y}$ lattice system, with the
nearest-neighbor hopping $t$. 
The spatial site index runs from $1$ to $L_{x}\times L_{y}$. 
We impose the periodic boundary condition. 
The superconducting gap equations is given 
as $\Delta_{ij} = V_{ij} \langle c_{i,\down} c_{j, \up} \rangle$, with 
pairing interaction $V_{ij}$. 
This equation is self-consistently solved by the polynomial expansion
method, as shown in Sec.~\ref{sec:polynomial_expansion}. 
The parameters in the SS method are set as $L_{0} = 10$, $M = 16$ and
$\kappa = 1.5$. 
We remove the eigen-pair $(\epsilon_{i},\bm{x}_{i})$
whose relative residual is greater than $10^{-1}$, as spurious
eigenpairs. 
See Eq.~(\ref{eq:rlt_res}). 
We do not use the iterative refinement of a subspace in the
following examples. 

\subsection{Computational costs in self-consistent calculations}
We use the polynomial expansion scheme to obtain self-consistent
superconducting gaps. 
Let us evaluate computational costs for a $d$-wave superconductor at
zero temperature, on a square lattice ($L_{x}=L_{y}$). 
An evaluation for an $s$-wave superconductor was shown in our previous
contribution\,\cite{NagaiJPSJ}. 
We measure the elapsed time for calculating
and updating the order parameters at one iteration.

\begin{figure}
\begin{center}
\resizebox{1 \columnwidth}{!}{\includegraphics{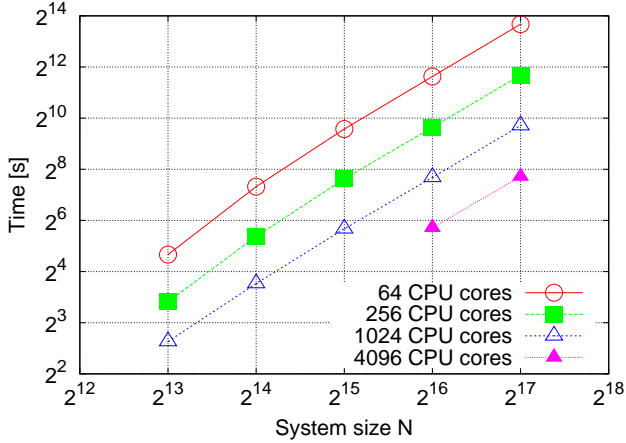}}
\end{center}
\caption{\label{fig:cpu} (Color online) System-size dependence of
 elapsed time for calculating and updating order-parameters at one 
 iteration with the polynomial expansion scheme for a $d$-wave
 superconductor at zero temperature, on a $L_{x}\times L_{y}$ square
 lattice ($L_{x}=L_{y}$). The system size is $N = L_{x} \times L_{y}$.} 
\end{figure}

\begin{figure}
\begin{center}
\resizebox{1 \columnwidth}{!}{\includegraphics{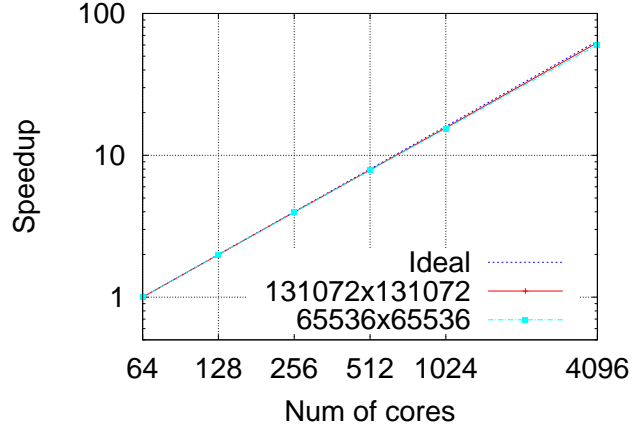}}
\end{center}
\caption{\label{fig:strong}(Color online) Strong-scaling plot in
 self-consistent calculations with the polynomial expansion scheme.}
\end{figure}

We used a supercomputing system PRIMERGY BX900 in Japan Atomic Energy
Agency. 
As shown in Fig.~\ref{fig:cpu}, the elapsed time for one iteration is
subjected to an $O(N^{2})$ rule, with increasing the system size 
$N = 2(L_{x} \times L_{y})$.  
This tendency is kept from 32 to 4096 CPU cores. 
In contrast, the full diagonalization scheme inevitably demands
$O(N^{3})$ costs in the core part of a calculation.  
This is a big advantage of the polynomial expansion scheme. 
Furthermore, we focus on the strong scaling, as shown in Fig.~\ref{fig:strong}. 
One can see an excellent strong scaling up to 4096 CPU cores. 

\subsection{Eigenvalues in a vortex lattice system in an $s$-wave superconductor}
\label{subsec:ev_vortex_lattice}
\begin{figure}
\begin{center}
\resizebox{1 \columnwidth}{!}{\includegraphics{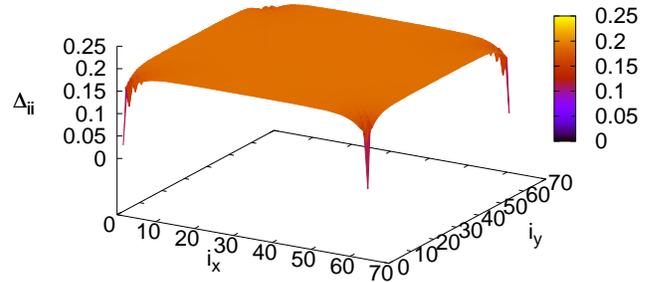}}
\end{center}
\caption{\label{fig:gap} (Color online) Order-parameter
 distribution obtained by a self-consistent calculation with the
 polynomial expansion scheme for a vortex lattice in an $s$-wave
 superconductor, with on-site hopping $V_{ii} = -1.5t$ and chemical
 potential $\mu = 0$. } 
\end{figure}

We show the eigenvalues obtained by the SS method.
The system in this section has a vortex square lattice in an $s$-wave
superconductor.  
The parameters are set as follows: on-site interaction 
$V_{ii} = -1.5t$, chemical potential $\mu = 0$, and spatial size 
$L_{x} \times L_{y} = 64 \times 64$. 
The matrix dimension is $N = 8192$ with 49152 nonzero entries. 
The rescaling parameters in the polynomial expansion method are $a=8t$
and $b=0$. 
Also, a cut-off parameter in the polynomial expansion
scheme\cite{NagaiJPSJ} is $2000$. 
The resultant order parameter is shown in Fig.~\ref{fig:gap}. 

The relative residual for the eigen-pair
\mbox{$(\epsilon_{i},\Vec{x}_{i})$} is calculated by  
\begin{align}
{\rm res}_{i} 
&= 
\frac{||H \Vec{x}_{i} - \epsilon_{i} \Vec{x}_{i}||}
{||H \Vec{x}_{i}|| + |\epsilon_{i}| ||\Vec{x}_{i}||}.
\label{eq:rlt_res}
\end{align}
After the self-consistent calculations with 100 iterations, we obtain
the eigen-pairs with the use of the SS method. 

The quadrature points are set by Eq.~(\ref{eq:daz}) and the
corresponding weights are set by Eq.~(\ref{eq:daw}), with 
$\alpha = 0.5$.  
The contour $\Gamma$ is set as $\gamma = 0$ and $\rho = 0.15t$. 
We adopt a sparse solver PARDISO\cite{PARDISO} to compute
Eq.~(\ref{eq:lin}). 
This solver uses the nested dissection algorithm from the METIS
package\cite{METIS}. 
We also confirm that the calculations with the shifted BiCG method, which
is suitable for large sparse matrices, have a similar result.  

\begin{figure}
\begin{center}
     \begin{tabular}{p{1 \columnwidth}} 
      \resizebox{1 \columnwidth}{!}{ \includegraphics{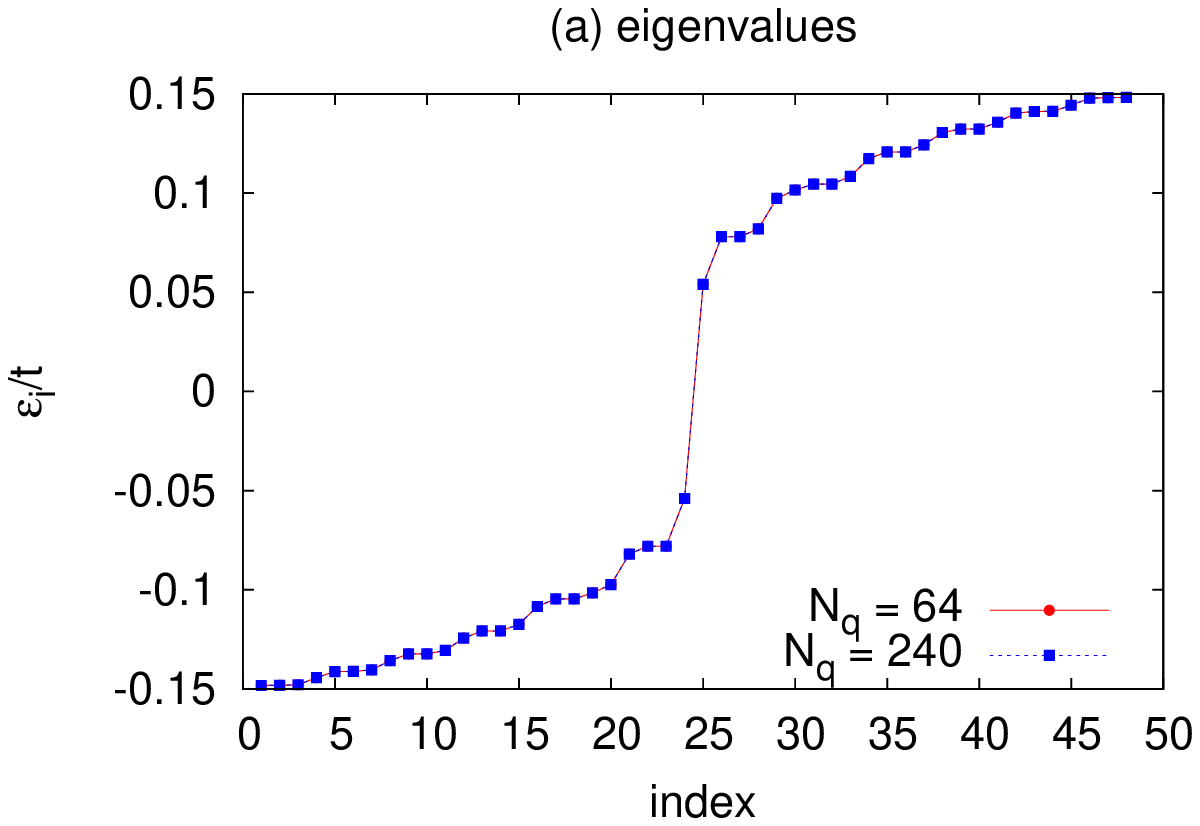}} \\
      \resizebox{1 \columnwidth}{!}{\includegraphics{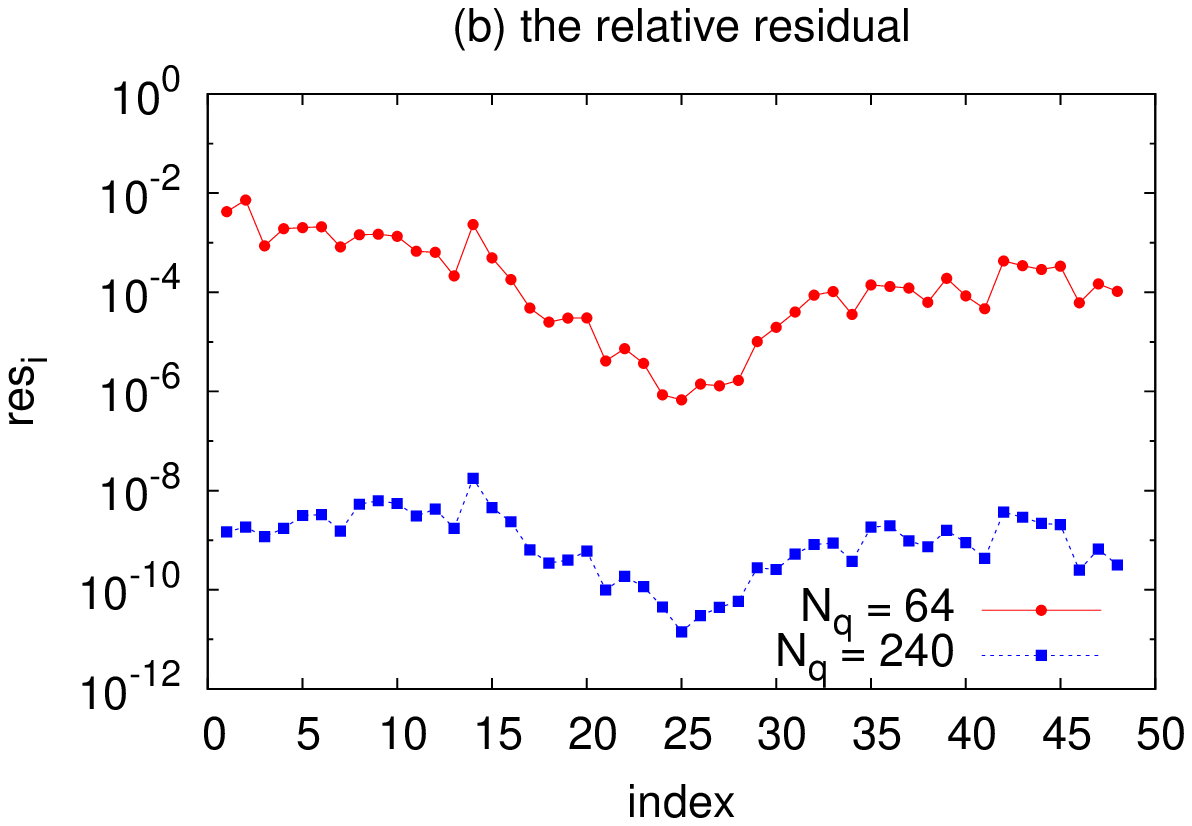}} 
    \end{tabular}
\end{center}
\caption{
\label{fig:qp}(Color online) (a) Eigenvalues of the
 Bogoliubov-de Gennes Hamiltonian for the order parameter shown in
 Fig.~\ref{fig:gap}, changing the total number of the quadrature points
 $N_{\rm q}$. The horizontal axis represents the index of the
 eigenvalue, in ascending order. 
(b) Relative residual for each eigen-pair
 \mbox{$(\epsilon_{i},\Vec{x}_{i})$}, varying $N_{\rm q}$. The index $i$
 in the horizontal axis is the same as in (a).}
\end{figure}

Let us investigate $N_{\rm q}$-dependence of the eigenvalues and the
relative residual. 
Figure \ref{fig:qp} shows that a calculation with $N_{\rm q}=64$
has good precision about the eigenvalues inside $\Gamma$ 
\mbox{($-0.15 < \epsilon_{i} < 0.15$)}. 
We note that the calculation with $N_{\rm q} = 64$ takes about 
{\it 30 seconds} with {\it only one} CPU core (Intel Xeon X5550 2.66GHz) by a
desktop computer.  
The conventional full diagonalization takes about 40 minutes with the same machine. 
When the system size becomes quadruple
\mbox{($L_{x} \times L_{y} = 128 \times 128$)}, it takes about 5 minutes
to obtain the same eigenvalue distribution, with the same one CPU core. 

\begin{figure}
\begin{center}
\resizebox{1 \columnwidth}{!}{\includegraphics{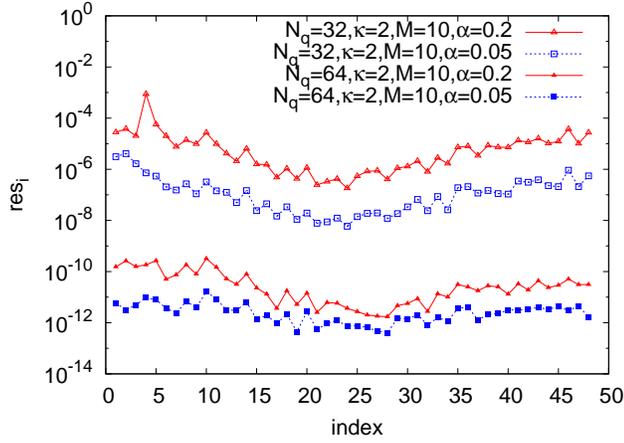}}
\end{center}
\caption{\label{fig:resipara}
(Color online) Accuracy of the Sakurai-Sugiura method in terms
 of relative residual (\ref{eq:rlt_res}). 
The horizontal axis corresponds to the index of the eigen-pair, as seen
 in Fig.~\ref{fig:qp}. 
The physical parameters are the same as in Fig.~\ref{fig:gap}, but the
 parameters related to the numerical calculations are changed.  
The total number of the quadrature points for contour integrals is set
 as either $N_{\rm q}=32$ or $64$. The source matrix size $L$ in
 Eq.~(\ref{eq:source_size}) is changed by varying a numeric constant
 $\kappa$ and the moment size $M$. The contour in the contour integrals
 is continuously deformed by the vertical scaling factor $\alpha$, as
 seen in Eq.~(\ref{eq:daz}).
}
\end{figure}

We discuss the accuracy of an eigenvalue calculation in the SS method. 
This issue depends on the parameters related to the contour integral
representation, as well as the source size $L$. 
The simplest improvement can be achieved by increasing the total number
of the quadrature points, $N_{\rm q}$. 
Alternatively, we obtain better accuracy with smaller $N_{\rm q}$,
continuously deforming the contour $\Gamma$. 
Figure ~\ref{fig:resipara} shows the relative residual, varying $N_{\rm q}$ and the
vertical scaling factor $\alpha$. 
We find that the accuracy becomes higher than Fig.~\ref{fig:qp}(b), even
though $N_{\rm q}$ is small. 
Here, we take a relatively larger source size ($\kappa=2$ and $M=10$ in
Eq.~(\ref{eq:source_size})), compared with the previous calculations.

\subsection{Magnetic-field dependence of the eigenvalue in long-coherence-length superconductors}
\label{subsec:ev_long_coherence_length}
We show the magnetic-field dependence of the
eigenvalues in a long-coherence-length superconductor.   
A coherence length of a superconductor $\xi$ is roughly estimated by the
ratio of the Fermi velocity to the amplitude of the order-parameter
($\xi \sim v_{F}/\Delta$). 
In many materials expect for high-$T_{c}$ cuprates, the coherence length
is much larger than an atomic length. 
Typically, the electric states in such superconducting systems are
described by the quasiclassical Eilenberger theory\cite{Eilenberger}, with
neglecting atomic-scale physics. 
However, interesting microscopic phenomena such as an interference
effect and discretized quantum bound states in a vortex core are never
treated in this approach. 
The mean-field BdG approach can treat these atomic-scale phenomena, but
requires extremely large computational costs when the coherence length is
much larger than an atomic length. 
Thus, to solve the BdG equations in a long-coherence-length
superconductor is a challenging issue.

\begin{figure}
\begin{center}
\resizebox{1 \columnwidth}{!}{\includegraphics{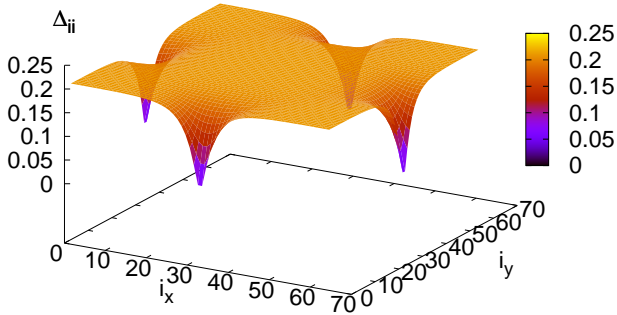}}
\end{center}
\caption{\label{fig:gaplong}(Color online) Order-parameter
 distribution obtained by a self-consistent calculation with the
 polynomial expansion scheme for a vortex lattice in an $s$-wave
 superconductor,  with on-site hopping $V_{ii} = -2t$, chemical
 potential $\mu = -t$, temperature $T = 0.04t$. 
The spatial size is given by $L_{x} = L_{y} = 64$.}
\end{figure}

\begin{figure}
\begin{center}
     \begin{tabular}{p{1 \columnwidth}} 
      \resizebox{1 \columnwidth}{!}{ \includegraphics{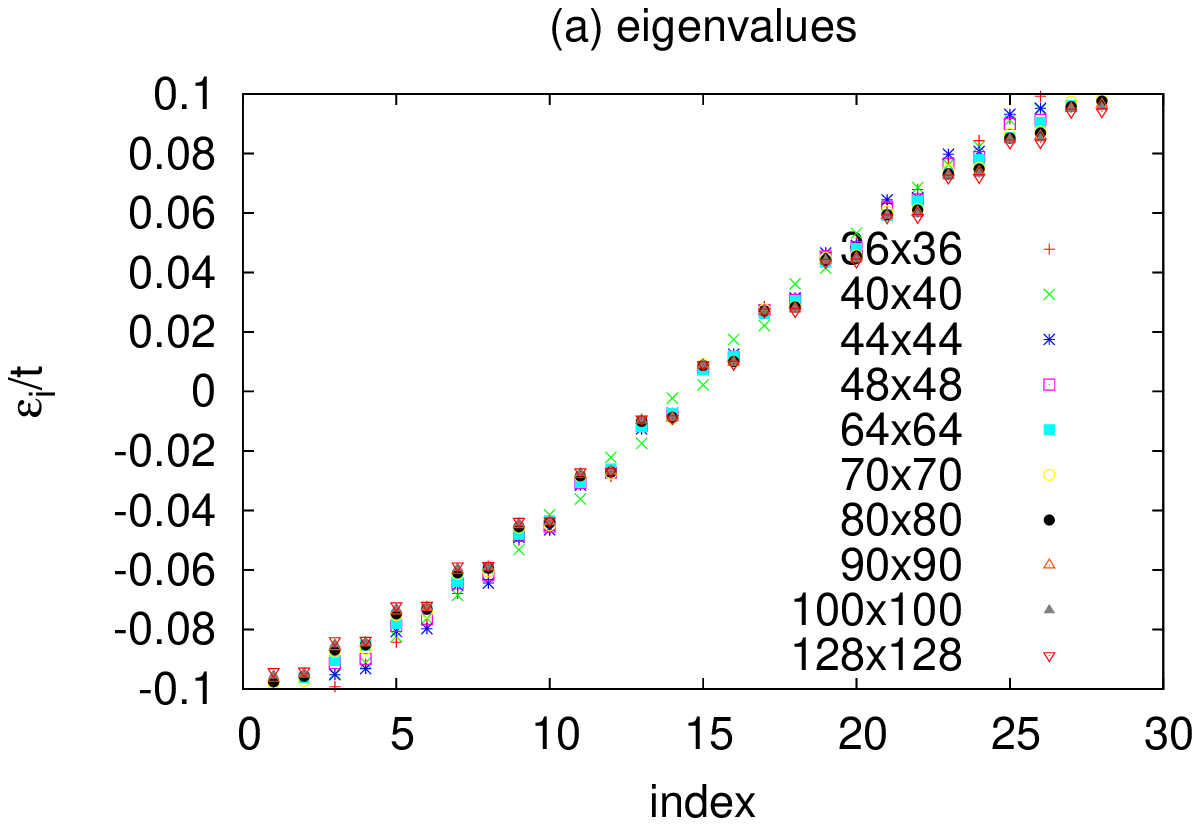}} \\
      \resizebox{1 \columnwidth}{!}{\includegraphics{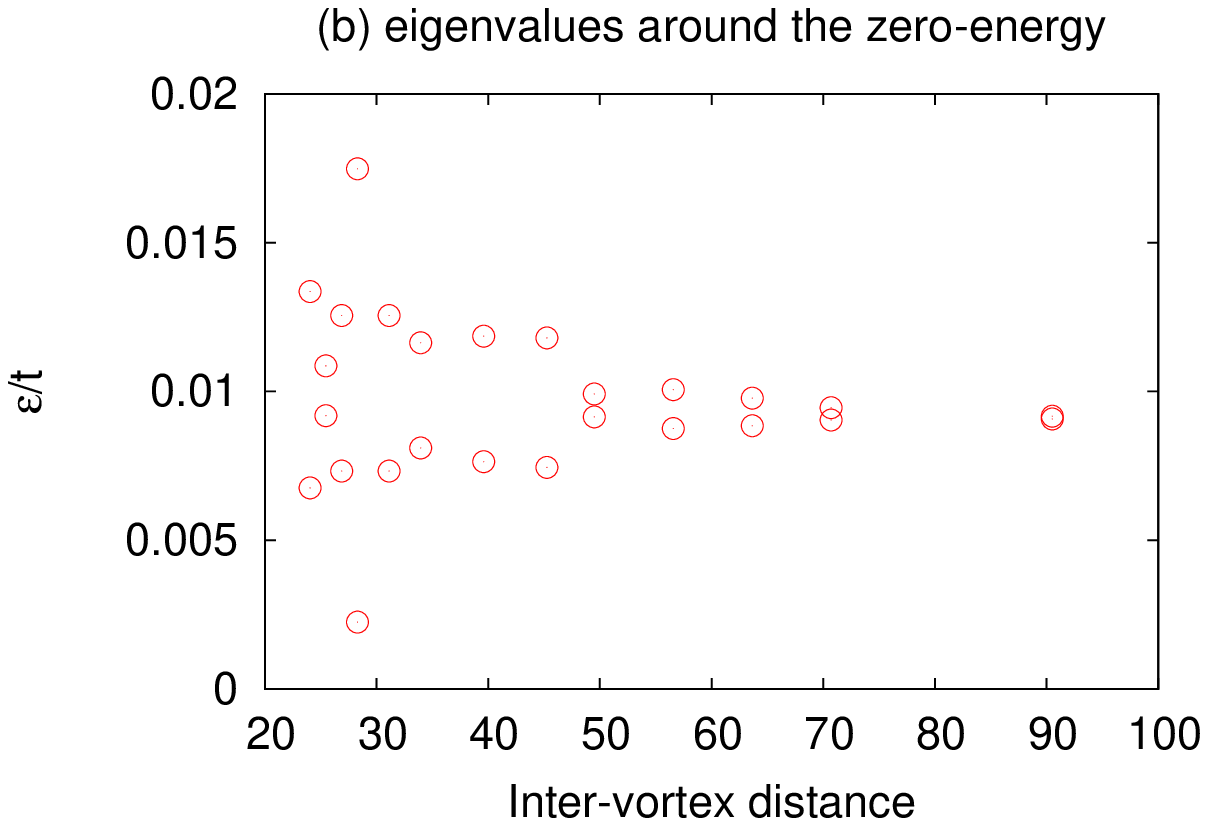}} 
    \end{tabular}
\end{center}
\caption{\label{fig:eigens}(Color online) Magnetic-field dependence of
 the distribution of the eigenvalues as functions of (a) the eigenvalue
 index and (b) the inter-vortex distance.} 
\end{figure}

We consider a vortex lattice in an $s$-wave
superconductor, with on-site interaction $V_{ii} = -2 t$ and chemical
potential $\mu = -t$. 
These parameters correspond to a model with a relatively smaller Fermi
surface, compared with Sec.~\ref{subsec:ev_vortex_lattice}. 
The temperature is set as $T = 0.04t$. 
We use the domain $\Gamma$ with $\gamma = 0$ and $\rho = 0.1t$ 
($-0.1t < \epsilon_{i} < 0.1t$), and $N_{\rm q} = 64$.  
The magnetic field becomes small with decreasing the system size, since the total magnetic flux is fixed. 
As shown in Fig.~\ref{fig:gaplong}, the amplitude of the order-parameter
is similar to that in the previous section.
Combined with the small Fermi surface, the coherence length is longer than in Sec.~VB. 
Under the periodic boundary condition, two bound states appear at each vortex core. 
Their energy eigenvalues are degenerate when the system size is large (low magnetic field).
With increasing the magnetic field, a splitting in the degenerate
eigenvalues occurs, as shown in  Fig,~\ref{fig:eigens}(a). 
The splitting comes from the occurrence of an overlap between the bound
states in a vortex core. 
Figure \ref{fig:eigens}(b) shows quantum oscillation as a function of
the inter-vortex distance originating from an interference effect
between two vortex bound states. 
We note that it is hard to discuss the degeneracy splitting with the only use of the polynomial expansion scheme, since the 
polynomial expansion calculates the local density of the states not the eigenvalues. 

\subsection{Magnetic field dependence of the thermal conductivity}
\begin{figure}
\begin{center}
\resizebox{0.8 \columnwidth}{!}{\includegraphics{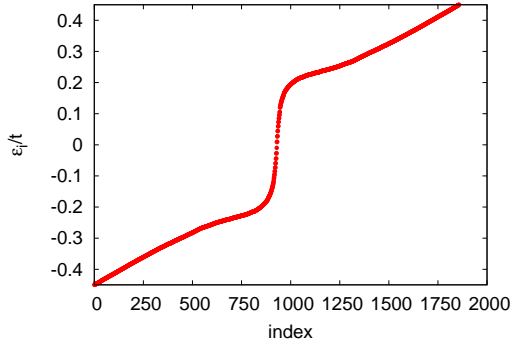}}
\end{center}
\caption{\label{fig:eigen90}(Color online) Eigenvalue distribution in an
 $s$-wave superconductor with a vortex lattice. The spatial size is
 given by $L_{x} = L_{y} = 90$.}
\end{figure}

We show the magnetic-field-dependence of the thermal conductivity in an
$s$-wave superconductor, with a vortex lattice. 
All the physical parameters are the same as the ones in
Sec.~\ref{subsec:ev_long_coherence_length}. 
We use the domain $\Gamma$ with $\gamma = 0$ and $\rho = 0.45t$ 
($-0.45t < \epsilon_{\alpha} < 0.45t$), and $N_{\rm q} = 64$. 
It takes about one hour to obtain 1857 eigenvalues located
in the domain $\Gamma$ with the same one CPU core when the system size
is $L_{x} \times L_{y} = 90 \times 90$ as shown in
Fig.~\ref{fig:eigen90}.  
One can clearly find that the gap amplitude in the bulk states is
$\Delta_{0} \sim 0.2t$, which is consistent with the estimation with the
use of Fig.~\ref{fig:gaplong}. 

Let us consider the case when a temperature gradient exists along
$x$-axis. 
Using the linear response theory\,\cite{Kubo;Nakajima:1957}, the
electronic thermal conductivity per volume of a superconductor is 
\begin{eqnarray}
 \kappa_{xx} 
&=& 
\frac{1}{T}
\lim_{\Omega \to 0}\frac{1}{\Omega} 
{\rm Im}\,
[P_{xx}(\imu \Omega_{m} \to \Omega + i0)]
\label{eq:def_thermal_conductivity} \\
&=&
\frac{1}{T}
\sum_{\gamma,\gamma^{\prime}}
F_{\gamma\gamma^{\prime}}
|
[\hat{U}^{\dagger}\hat{V}^{(x)}\hat{U}]_{\gamma^{\prime}\gamma}
|^{2}, 
\end{eqnarray} 
with
\begin{eqnarray}
F_{\gamma\gamma^{\prime}}
&=&
\int\frac{d\omega}{2\pi}
\delta_{\eta}(\omega-\epsilon_{\gamma}) 
\int\frac{d\omega^{\prime}}{2\pi}
\delta_{\eta}(\omega^{\prime}-\epsilon_{\gamma^{\prime}}) 
\nonumber \\
&&
\times \pi \delta(\omega-\omega^{\prime})
\omega^{2}f^{\prime}(\omega). 
\label{eq:def_F_gammagamma}
\end{eqnarray} 
Here, $P_{xx}(\imu \Omega_{m})$ is the Fourier transformation of a
current-current correlation
function\,\cite{Langer:1962,Ambegaokar;Teword:1964,Takigawa;Machida:2002}
in the imaginary time, 
\(
 P_{xx}(\tau,0)
=
\langle
T_{\tau} [J_{{\rm u},x}(\tau)J_{{\rm u},x}(0)]
\rangle
\), where $J_{{\rm u},x}$ is the Heisenberg operator of energy
flux\cite{Langer:1962} along $x$-axis. 
The Lorentian kernel 
\mbox{
$\delta_{\eta}(\omega)=\eta/[\pi (\omega^{2}+\eta^{2})]$
} in $F_{\gamma\gamma^{\prime}}$ represents a dissipation
effect\,\cite{Takigawa;Machida:2002}.  
The matrix $\hat{U}$ contains the eigenvectors of the BdG Hamiltonian,
while the matrix $\hat{V}^{(x)}$ includes contributions from the energy
flux. 
We show their explicit formulae in Appendix
\ref{app:thermal_conductivity}.  

\begin{figure}
\begin{center}
     \begin{tabular}{p{1 \columnwidth}} 
      \resizebox{1 \columnwidth}{!}{ \includegraphics{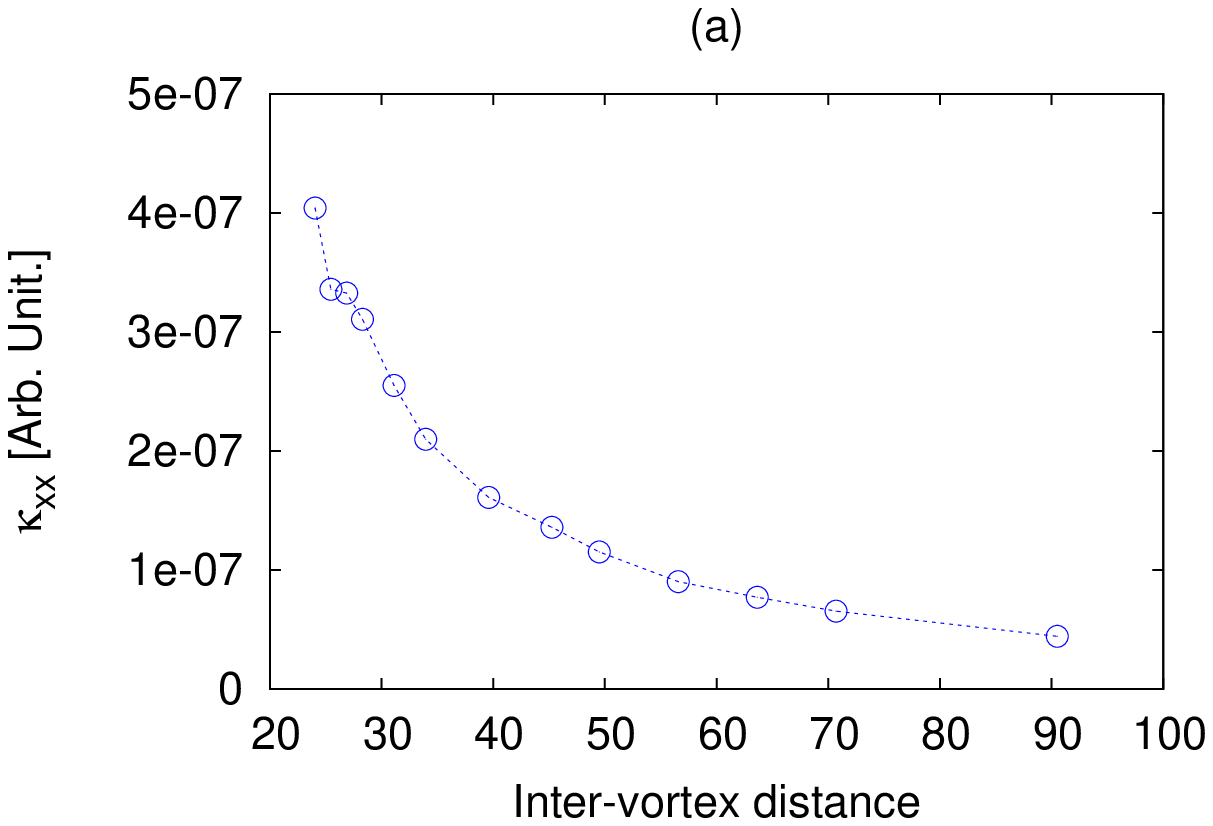}} \\
      \resizebox{1 \columnwidth}{!}{\includegraphics{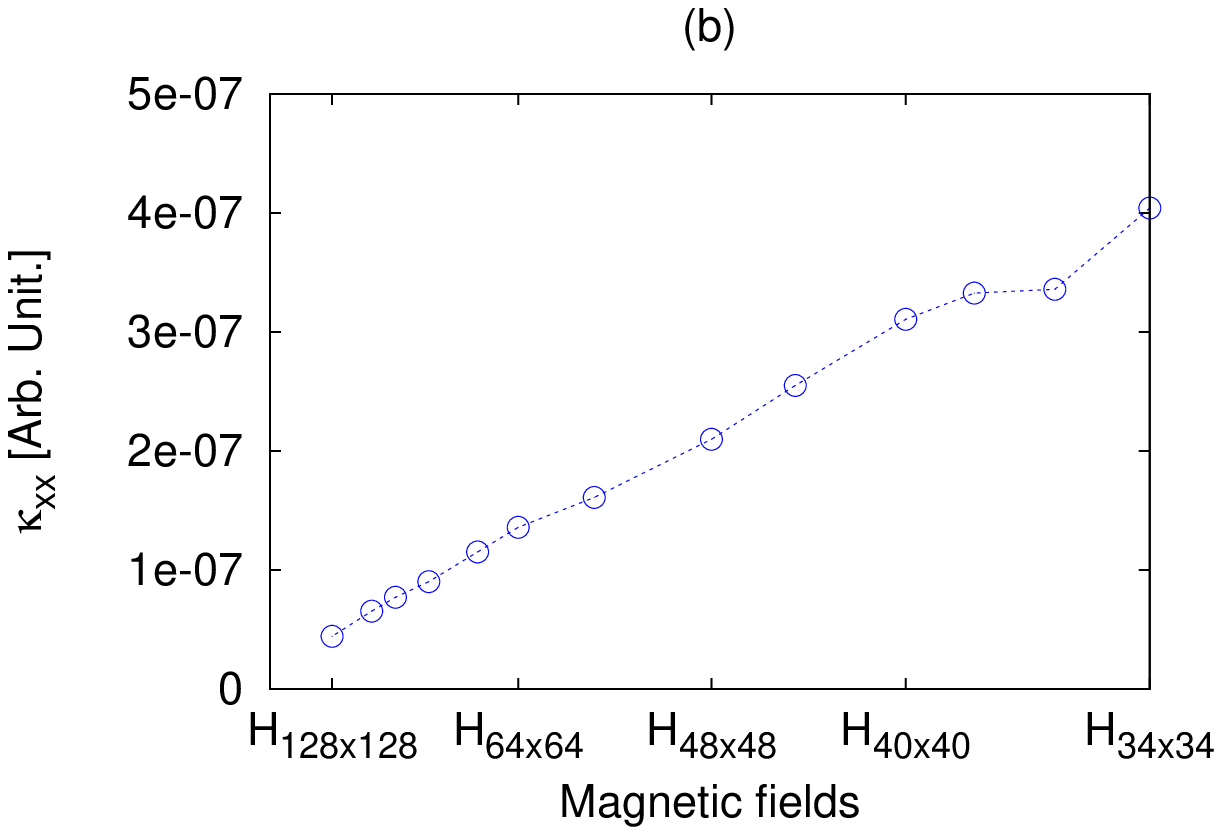}} 
    \end{tabular}
\end{center}
\caption{\label{fig:kappa}(Color online) Magnetic-field dependence of
 the thermal conductivity as a function of (a) an inter-vortex distance and (b) magnetic fields. The inter-vortex distance become long, when decreasing magnetic field.} 
\end{figure}

We adopt the damping factor $\eta = 0.005t$. 
We assume that $\theta_{i j} = 0$ in Eq.~(\ref{eq:def_energy_flow})
because the vector potential around a vortex is small. 
Figure \ref{fig:kappa} shows that $\kappa_{xx}$ drastically increases when the inter-vortex distance is shorter than 
around 60 (i.e., a high magnetic-field domain). 
This behavior could be related to an interference effect between the two bound states in vortex cores in high magnetic field, as seen 
Fig.~\ref{fig:gaplong}. 
A quantum oscillation in the eigenvalue distribution becomes remarkable in such a high magnetic-field domain.

\subsection{Temperature dependence of nuclear magnetic relaxation rates}
We show the temperature dependence of the nuclear magnetic relaxation rate
in a uniform $s$-wave superconductor. 
It is well known that the nuclear magnetic relaxation rate
$T_{1}(\Vec{r}_{i},T)$ is calculated by\cite{Takigawa} 
\begin{align}
T_{1}(\Vec{r}_{i},T) &= \frac{1}{R(\Vec{r}_{i},T)}, \\
R(\Vec{r}_{i},T) &= \lim_{\Omega \rightarrow 0^{+}} {\rm Im} \: \frac{\chi_{-+}(\Vec{r_{i}},\Vec{r}_{i}; \Omega)}{\Omega/T}, \\
&= - \sum_{\alpha \beta,\epsilon_{\alpha}>0,\epsilon_{\beta} >0} U_{i \alpha}U_{i \beta}^{\ast} \left[ 
U_{i \alpha} U_{i \beta}^{\ast} +U_{i+N \alpha} U_{i+N \beta}^{\ast} 
\right] 
\nonumber \\
& \times 
\pi T f'(\epsilon_{\alpha}) \delta(\epsilon_{\alpha} - \epsilon_{\beta}).
\end{align}
We use the eigenvalues in the domain with $\gamma = 0.25t$, $\rho = 0.25t$ ($0 < \epsilon_{\alpha} < 0.5t$). 
The parameters are set as follows: the onsite interaction $V_{ii} = - 2t$, the chemical potential $\mu = -t$, and the system size $L_{x} \times L_{y} = 64 \times 64$. 
The delta function $\delta(x)$ is approximated by $\delta(x) = (1/\pi) \eta/(x^{2} + \eta^{2})$ with the smearing factor $\eta = 0.01t$.  
We adopt the shifted BiCG method as a iterative linear solver.
\begin{figure}
\begin{center}
\resizebox{0.8 \columnwidth}{!}{\includegraphics{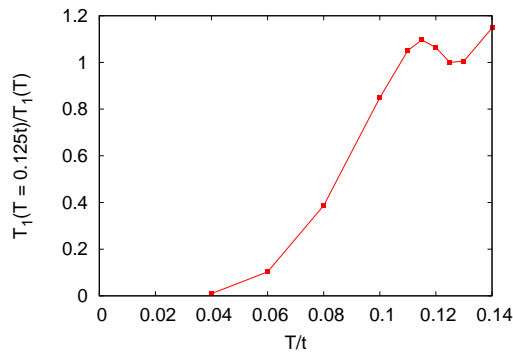}}
\end{center}
\caption{\label{fig:nmr}(Color online) Temperature dependence of the
 nuclear magnetic relaxation rate in a uniform $s$-wave
 superconductor. $L_{x} = L_{y} = 64$. }
\end{figure}
As shown in Fig.~\ref{fig:nmr}, the nuclear magnetic relaxation rate can
be successfully reproduced by the SS method.  
We note that the discrete energy levels due to the finite size system
cause the relatively smaller Hebel-Slichter peak below $T_{c}$.  
We mention here that the accuracy of calculating physical quantities by the SS method depends on the
size of an energy domain on the complex plain. 
A truncation error may occur in evaluating Eq.~(\ref{eq:dynam}), if this size is not enough large. 
Figure \ref{fig:resipara} indicates the eigen-pairs inside $\Gamma$ are evaluated with high accuracy.
Therefore, the accuracy of evaluating the nuclear magnetic relaxation rate around $T_c$ may increase, taking a relatively
larger domain size. 

%
\subsection{Computational costs in the eigenvalue problem with the SS-method in a vortex lattice system}
\begin{figure}
\begin{center}
\resizebox{0.6 \columnwidth}{!}{\includegraphics{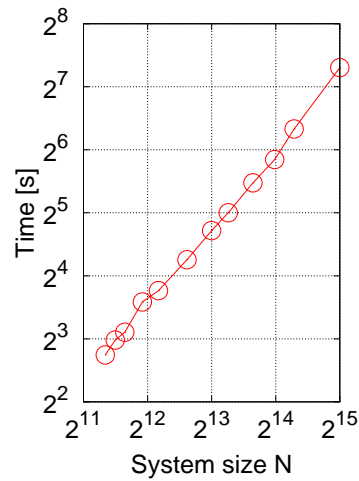}}
\end{center}
\caption{\label{fig:tdepSS}(Color online) System-size dependence of the
 elapsed time with one CPU core for obtaining the eigenvalues 
$(-0.1t < \epsilon_{i} < 0.1t)$ with the SS-method, for an $s$-wave
 superconductor at $T = 0.04t$, on a $L_{x}\times L_{y}$ square lattice
 ($L_{x}=L_{y}$). The system size is $N = 2 L_{x}\times L_{y}$.}
\end{figure}

Now, let us evaluate the computational costs of
the SS method. 
We measure the elapsed time from reading the Hamiltonian matrix
constructed by the polynomial expansion scheme to  
finishing the SS-method in a $L_{x} \times L_{y}$ square lattice
$s$-wave superconductor at $T = 0.04t$ ($L_{x}=L_{y}$).  
We use the contour $\Gamma$ with $\gamma = 0$ and $\rho = 0.1t$, and the
physical parameters are the same as in Fig.~\ref{fig:eigens}.  
For the measurement, we use a desktop computer with only one CPU core
(Intel Xeon X5550 2.66GHz). 
As shown in Fig.~\ref{fig:tdepSS}, the elapsed time of the SS method
grows in an ${\cal O}(N)$ manner with increasing the system size 
$N = 2(L_{x} \times L_{y})$.   
The computational costs are roughly estimated by ${\cal O}(m_{\rm s} N)$. 
In all the calculation of this subsection,
the energy domain is restricted to
$-0.1t < \epsilon_{\alpha} < 0.1t$.
As a result, the number of the eigenvalues $m_{s}$ is independent of the spatial size $N$.
Indeed, we find that only the bound states in vortices are relevant to this narrow energy window. 
If an energy domain is wide (large $\rho$), $m_{s}$ increases. 
In this case, the computational cost predominantly depends on the Gram-Schmidt orthonormalization procedure 
to 
construct $\tilde{Q}$. 
Then, we find that the cost is estimated by ${\cal O}(N m_{\rm s}^{2})$. 
When one tries to obtain all the eigenvalues (i.e., $m_{s} = N$) with a wide energy domain, the cost of the SS method 
becomes ${\cal O}(N^{3})$. 
This result is equivalent to the one in the full diagonalization method. 
However, the wide energy domain is easily divided into small energy domains. 
For example, with using $N_{\rm s}$ domains with $N_{\rm s}$ parallel computation
with $N_{\rm s}$ CPU cores, the elapsed time reduces to $1/N_{\rm s}$.  
This is a big advantage of the SS method.      
\section{Conclusion}
\label{sec:conclusion}
We proposed the fast efficient method on the basis of the SS-method and
the polynomial expansion scheme to  
calculate the eigen-pairs and the dynamical correlation functions in the BdG scheme of superconductivity. 
The polynomial expansion scheme enables us to solve the gap-equations self consistently, with large scale parallel computations. 
With the use of the SS method, one can solve
issues for finding eigenvalues in a given energy domain 
 and their corresponding eigenvectors. 
The virtue of the SS method is to reduce systematically
the size of a large Hamiltonian, keeping its predominant
contributions in a low-energy scale.
In other words, this proposal leads to a numerical
construction of an effective low-energy Hamiltonian
in the BdG approach of superconductivity.
We applied the present approach to the calculations of various physical quantities including the eigenvalues distribution of the BdG Hamiltonian in a vortex lattice, magnetic-field-dependence of the thermal conductivity, and temperature-dependence of the nuclear magnetic relaxation rate. 
We stress that most of the calculations were performed, changing the system size. 
This is quite important for developing a theoretical tool to predict physical behaviors in nano-scale superconductors from a microscopic theory. 

\begin{acknowledgment}
The authors would like to acknowledge Masahiko Machida and Susumu Yamada for helpful discussions and comments. 
The calculations have been performed using the supercomputing 
system PRIMERGY BX900 at the Japan Atomic Energy Agency. 
This study has been supported by Grants-in-Aid for Scientific Research from MEXT of Japan. 
Y.O. is supported in part by the Special Postdoctoral
Researchers Program, RIKEN.
\end{acknowledgment}

\appendix
\section{Iterative refinement}
\label{app:iterative_refinement}
In Sec.~\ref{subsec:construction_subspace}, we remark that 
larger subspace makes more higher accuracy but needs more heavier computational costs. 
A problem may occur when we choose a small value of $L$ to avoid the use of a large subspace. 
If some residuals of the obtained approximate eigenvalues and
eigenvectors are not small enough for a given tolerance, we can brush up
the resulting approximate eigenvalues and eigenvectors. 
We propose two ways to brush up the eigen pairs with the choice of the appropriate 
source matrix $\hat{V}$ dominantly constructed by the information in a given domain $\Gamma$. 

First, one can use the source matrix $\hat{V}$ as 
\begin{align}
\hat{V} = \hat{S}_{0}, \label{eq:vv}
\end{align}
Then, we implement $r$ iterations of $P_{\Gamma}(A)$ on $\hat{V}$, 
\begin{equation}
\hat{S}_{0}^{(r)} = P_{\Gamma}(A)\hat{S}_{0}^{(r-1)}, 
\quad
\hat{S}_{0}^{(0)} = \hat{V}.
\end{equation}
Using $\hat{S}_{0}^{(r-1)}$, we construct a refined matrix including a
higher moment vector, 
\mbox{
\(
\hat{S}_{k}^{(r)} = A^{k} P_{\Gamma}(A) \hat{S}_{0}^{(r-1)}
\)}. 
In a simulation, numerical quadrature is used for calculating
$P_{\Gamma}(A)$ and $A^{k}P_{\Gamma}(A)$, as seen in
Sec.~\ref{subsec:numerical_quadrature}. 
Thus, performing the singular-value decomposition of 
$\hat{S}^{(r)}=\{\hat{S}_{0}^{(r)},\ldots,\hat{S}_{M-1}^{(r)}\}$, we
evaluate a refined effective rank $m_{\rm s}$. 

Second, we can brush up
the resulting approximate eigenvalues and eigenvectors by setting the
source matrix as  
\begin{align}
\hat{V} = \{\Vec{x}_{1},\cdots, \Vec{x}_{m_{\rm s}} \} \hat{C},\label{eq:vc}
\end{align}
where $\hat{C} \in \mathbb{C}^{m_{\rm s} \times L}$ whose elements are
random numbers in $(-1,1)$, and $\Vec{x}_{1},\cdots,\Vec{x}_{m_{\rm s}}$ are
the selected eigenvectors that are regarded as the approximate
eigenvectors with respect to the eigenvalues inside $\Gamma$. 
Using this $\hat{V}$, we reevaluate $\hat{S}$ and $m_{\rm s}$ in
Sec.~\ref{subsec:numerical_quadrature}. 

\section{Estimation of the trace}
\label{app:estimation_trace}
We show that the trace of an $n \times n$ matrix $A$ can be estimated by 
\begin{align}
{\rm Tr} \: A &\sim \frac{1}{s} \sum_{k=1}^{s} (\Vec{v}^{k})^{\rm T}  A \Vec{v}^{k},
\end{align}
with random vectors $\Vec{v}^{k}$ with entries $\pm 1$. 
If the vectors $\Vec{v}_{k}$ have entries $\pm 1$, the right-hand side
in the above equation is expressed as 
\begin{align}
\frac{1}{s} \sum_{k=1}^{s} (\Vec{v}^{k})^{\rm T} A \Vec{v}^{k} 
&= {\rm Tr} \:  A + 
\frac{1}{s} \sum_{ij,i \neq j}^{n} A_{ij} \sum_{k=1}^{s}  (\Vec{v}^{k})_{i}  (\Vec{v}^{k})_{j}. 
\end{align}
On the average, the coefficient of $A_{ij}$ in the above expansion will converge to zero provided that the components of 
the vectors $\Vec{v}^{k}$ have balanced $\pm$ signs. 
 
\section{Thermal conductivity}
\label{app:thermal_conductivity}
We derive the expression of the thermal conductivity in terms of the
solution of the BdG equation.  
Let us consider the case when a temperature gradient exists along
$x$-axis on a two-dimensional $L_{x}\times L_{y}$ lattice with lattice constant $a$.   
The electronic thermal conductivity per volume of a superconductor
associated with heat flux along $x$-axis is given in
Eq.~(\ref{eq:def_thermal_conductivity}). 
The current-current correlation function with respect to the energy flux
is 
\begin{equation}
 P_{xx}(\tau,0)
\equiv 
\langle
T_{\tau} [J_{{\rm u},x}(\tau)J_{{\rm u},x}(0)]
\rangle
=
\frac{1}{\beta}
\sum_{\imu \Omega_{m}}
e^{-\imu \Omega_{m}\tau}
P_{xx}(\imu \Omega_{m}). 
\end{equation}
The Heisenberg operator of energy flux\,\cite{Langer:1962} along
$x$-axis is  
\begin{equation}
 J_{{\rm u},x}
=
 \sum_{i,j}
\frac{1}{\imu}
\sum_{\sigma=\uparrow,\downarrow}
\left(
\frac{\partial c_{i,\sigma}^{\dagger}}{\partial \tau}
D_{x,ij}
c_{\sigma,j}
-
D_{x,ij}^{\ast}c_{j,\sigma}^{\dagger}
\frac{\partial c_{i,\sigma}}{\partial \tau}
 \right),
\label{eq:def_energy_flow}
\end{equation}
with 
\mbox{
\(
D_{x,ij} = \delta_{i+1_{x},j} (-iat)  e^{\imu \theta_{ij}}
\)} 
and
\mbox{ 
\(
1_{x}=(1,0)=\bm{a}/a
\)}. 
The link variable $ e^{\imu  \theta_{ij}}$ represents the contribution of
the magnetic field, 
\mbox{
\(
\theta_{ij} = (\pi/\phi_{0})\bm{a}\cdot \bm{A}[(\bm{r}_{i}+\bm{r}_{j})/2]
\)},
with the flux quantum $\phi_{0}$ and the vector potential
$\bm{A}$. 
The matrix \mbox{$D_{x}(=(D_{x,ij}))$} is related to the momentum operator on a
square lattice\,\cite{Takigawa;Machida:2002}. 

This current-current correlation function may be rewritten as the form
of a two-particle Green's function\,\cite{Ambegaokar;Teword:1964}, 
\begin{eqnarray}
P_{xx}(\tau_{1},\tau_{2}) 
&=&
\sum_{a,b,c,d}
\hat{J}^{(x)}_{ab}(\partial_{\tau_{1}^{\prime}},\partial_{\tau_{1}})
\hat{J}^{(x)}_{cd}(\partial_{\tau_{2}^{\prime}},\partial_{\tau_{2}})
\nonumber \\
&&
\times
\langle
T_{\tau}[
\psi_{a}^{\dagger}(\tau_{1}^{\prime})
\psi_{b}(\tau_{1})
\psi_{c}^{\dagger}(\tau_{2}^{\prime})
\psi_{d}(\tau_{2})
]
\rangle,
\end{eqnarray}
with 
\mbox{$\tau_{1}^{\prime}\to\tau_{1}+0$} and 
\mbox{$\tau_{2}^{\prime}\to\tau_{2}+0$}. 
We neglected some terms associated with the action of the
imaginary-time derivative on the time ordering operator, according to
the discussion by Ambegaokar and Teword\,\cite{Ambegaokar;Teword:1964}. 
Here, $\psi$ and $\psi^{\dagger}$ are defined as, respectively, 
\mbox{
\(
\psi
=(c_{\uparrow}, \bar{c}_{\downarrow})^{\rm T}
\)} 
and 
\mbox{
\(
\psi^{\dagger}
=(\bar{c}_{\uparrow}^{\rm T}, c_{\downarrow}^{\rm T})
\)}, with 
\mbox{
\(
c_{\sigma}=(c_{1,\sigma},\ldots)^{\rm T}
\)} 
and 
\mbox{
\(
\bar{c}_{\sigma}=(c^{\dagger}_{1,\sigma},\ldots)^{\rm T}
\)}. 
This convention corresponds to the Nambu representation. 
The indices $a$, $b$, $c$, and $d$ run from $1$ to $2L_{x}L_{y}$. 
The $2L_{x}L_{y}\times 2L_{x}L_{y}$ matrix \mbox{$\hat{J}^{(x)}$} is
defined as 
\begin{equation}
\hat{J}^{(x)}
=
\left(
\begin{array}{cc}
J^{(x)} &  0 \\
0 & -J^{(x)\,\ast}
\end{array}
\right), 
\end{equation}
with the $L_{x}L_{y}\times L_{x}L_{y}$ matrix 
\mbox{ 
\(
J^{(x)}(\partial_{\tau^{\prime}},\partial_{\tau})
=
-\imu (
\partial_{\tau^{\prime}}D_{x}
-
\partial_{\tau}D_{x}^{\dagger})
\)}. 
The formulation in Sec.\,\ref{sec:formulation}, which is developed in a more
general representation for a fermion mean-field theory, is
straightforwardly rewritten in terms of the Nambu representation. 

Now. let us derive the expression of the thermal conductivity. 
First, we evaluate the imaginary part of 
\mbox{$P_{xx}(\imu \Omega_{m}\to \Omega+i0)$}. 
Next, we expand the resultant formula up to $\Omega$, to take the limit
$\Omega \to 0$. 
We note that ${\rm Im}[P_{xx}(\Omega=0)]=0$. 
Then, we obtain Eq.~(\ref{eq:def_thermal_conductivity}). 
The $2L_{x}L_{y}\times 2L_{x}L_{y}$ matrix $\hat{V}^{(x)}$ in
Eq.~(\ref{eq:def_thermal_conductivity}) is defined as 
\begin{equation}
 \hat{V}^{(x)}
=
\left(
\begin{array}{cc}
D_{x}+D_{x}^{\dagger} & 0\\
0 & (D_{x}+D_{x}^{\dagger})^{\ast}
\end{array}
\right). 
\end{equation}



\end{document}